\documentclass[slac_one]{revtex4}
\usepackage{graphicx}
\usepackage{fancyhdr}
\usepackage{rotating}
\usepackage{amsmath}
\newcommand{\Ds}{\Delta m^2_{\textsc{s}}}
\newcommand{\Da}{\Delta m^2_{\textsc{a}}}

\begin{document}

%\title{LaTex Template for ICHEP08} %% Paper title goes here
%\title{Probing minimal supergravity in type-I seesaw with lepton flavour violation at the LHC} %% Paper title goes here
\title{Correlations between lepton flavour violation at the LHC and type-I seesaw parameters in minimal supergravity} %% Paper title goes here


\author{A. Villanova del Moral}
\email{albert@cftp.ist.utl.pt}
\affiliation{Departamento de F\'\i sica and CFTP, Instituto Superior T\'ecnico, Avenida Rovisco Pais 1, %$\:\:$ 
1049-001 Lisboa, Portugal}

\begin{abstract}
  The most general supersymmetric seesaw mechanism has too many
  parameters to be predictive and thus can not be excluded by {\em
    any} measurements of lepton flavour violating (LFV) processes. We
  focus on the simplest version of the type-I seesaw mechanism
  assuming minimal supergravity (mSugra) boundary conditions. We 
  compute branching ratios for the LFV scalar tau decays, ${\tilde
    \tau}_2 \to (e,\mu) + \chi^0_1$, as well as loop-induced LFV
  decays at low energy, such as $l_i \to l_j + \gamma$ and $l_i \to 3
  l_j$, exploring their sensitivity to the unknown seesaw parameters.
  We find some simple, extreme scenarios for the unknown right-handed
  parameters, where ratios of LFV branching ratios correlate with
  neutrino oscillation parameters. If the overall mass scale
  of the left neutrinos and the value of the reactor angle were known,
  the study of LFV allows, in principle, to extract information about
  the so far unknown right-handed neutrino parameters.
\end{abstract}

\maketitle

\thispagestyle{fancy}


%%%%%%%%%%%%%%%%%%%%%%%%%%%%%%%%%%%%%%%%%%%%%%%%%%%%%%%%%%%%%%%%%%%%%%%%%%%%%%%%%%%%%%%%%%%%%%%%%%%%%%%%%%%%%%%%%
\section{INTRODUCTION}
Neutrino experiments~\cite{Fukuda:1998mi} %\cite{Fukuda:1998mi,Ahmad:2002jz,Abe:2008ee,Arpesella:2008mt,Aharmim:2008kc,Adamson:2008zt} 
have firmly established that neutrinos are massive and global fits to all neutrino oscillation data~\cite{Maltoni:2004ei} %\cite{Maltoni:2004ei,Schwetz:2008er} 
give precise values for the neutrino mass splittings and their mixing angles. The evidence of massive neutrinos provides the first experimental signal of physics beyond the Standard Model, being the so-called seesaw mechanism~\cite{Minkowski:1977sc} the most popular mechanism to generate Majorana neutrino masses. %\cite{Minkowski:1977sc,GellMann:1980vs,Yanagida:1979as,Mohapatra:1979ia,Schechter:1980gr,Schechter:1981cv}.

In this work~\cite{Hirsch:2008dy} we study the correlation with neutrino parameters of the ratio of LFV decays of the stau, %Br(${\tilde\tau}_2 \to e + \chi^0_1$)/Br(${\tilde\tau}_2 \to \mu + \chi^0_1$)
assuming mSugra boundary conditions and the simplest type-I seesaw mechanism as the origin of neutrino masses and mixings. In this theoretical framework, left-slepton LFV decays are proportional to the square of the off-diagonal elements of the slepton mass matrix, due to the renormalization group equation (RGE) running of the soft breaking parameters.
%%%%%%%%%%%%%%%%%%%%%%%%%%%%%%%%%%%%%%%%%%%%%%%%%%%%%%%%%%%%%%%%%%%%%%%%%%%%%%%%%%%%%%%%%%%%%%%%%%%%%%%%%%%%%%%%%
\section{ANALYSIS}
\label{sec:analysis}
For qualitative understanding, we consider the leading-log approximated solutions to the RGE's. The left-slepton LFV decays are then proportional to
\begin{equation}
\textrm{Br}(\tilde l_i\to l_j\chi_1^0)\propto \left|(\Delta M_{\tilde L}^2)_{ij}\right|^2 \propto \left|(Y_{\nu}^{\dagger}LY_{\nu})_{ij}\right|^2. 
\end{equation}
We can parametrize the neutrino Yukawa matrix as~\cite{Casas:2001sr}
\begin{equation}\label{Ynu}
Y_{\nu} =\sqrt{2}\frac{i}{v_U}\sqrt{\hat M_R}R\sqrt{{\hat m_{\nu}}}U^{\dagger},
\end{equation}
where $\hat m_{\nu}$ and $\hat M_R$ are the diagonal matrices with 
the light neutrino mass eigenvalues $m_i$ and the heavy neutrino 
mass eigenvalues $M_i$, respectively; $U$ is the leptonic mixing matrix and $R$ is a complex orthogonal matrix. 
This way, the left-slepton LFV decays are correlated to neutrino parameters
\begin{equation}
\textrm{Br}(\tilde l_i\to l_j\chi_1^0)\propto \left|U_{i\alpha}U_{j\beta}^*\sqrt{m_{\alpha}}\sqrt{m_{\beta}}
R_{k\alpha}^*R_{k\beta}M_k\log\left(\frac{M_X}{M_k}\right)\right|^2. 
\end{equation}
In order to eliminate most of the dependence on the supersymmetric parameters, we work with ratios of LFV decays. 
Thus, for example, the ratio of stau LFV decays can be expressed in terms of the parameter $r^{13}_{23}$, 
\begin{equation}\label{eq:3}
\frac{\textrm{Br}({\tilde\tau}_2 \to  e +\chi^0_1)}
     {\textrm{Br}({\tilde\tau}_2 \to  \mu +\chi^0_1)}
 \simeq \frac{|(\Delta M_{\tilde L}^2)_{13}|^2}{|(\Delta M_{\tilde L}^2)_{23}|^2}
 \equiv \left(r^{13}_{23}\right)^2,
\end{equation}
which only depends on neutrino parameters.

\begin{figure}[htb]
\centering
\begin{tabular}{|c|c|c|c|c|}\cline{2-5}%\hline%\vspace{0.1cm}
%\hspace{0.6cm}\B 
\multicolumn{1}{c|}{}& {\bf SNH} & {\bf SIH} &  {\bf QDNH} & {\bf QDIH} \\\hline
\hspace{0.15cm}
\begin{rotate}{90}{$\qquad\ \ \boldsymbol{\delta=0}$}\end{rotate} &
\includegraphics[width=0.23\textwidth]{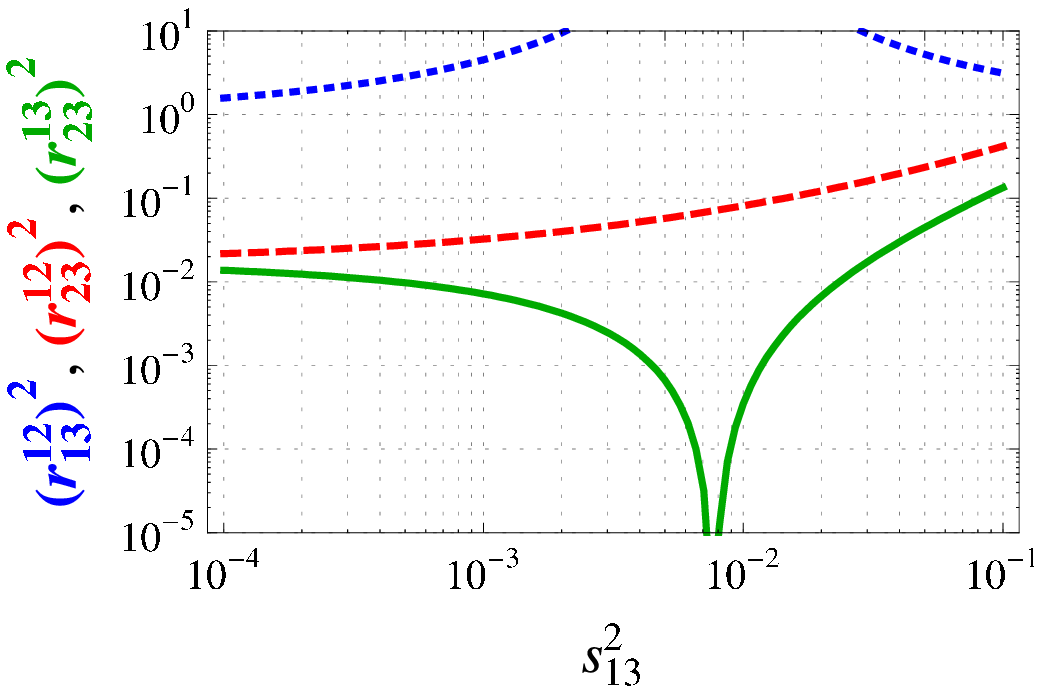} & % 0.38
\includegraphics[width=0.23\textwidth]{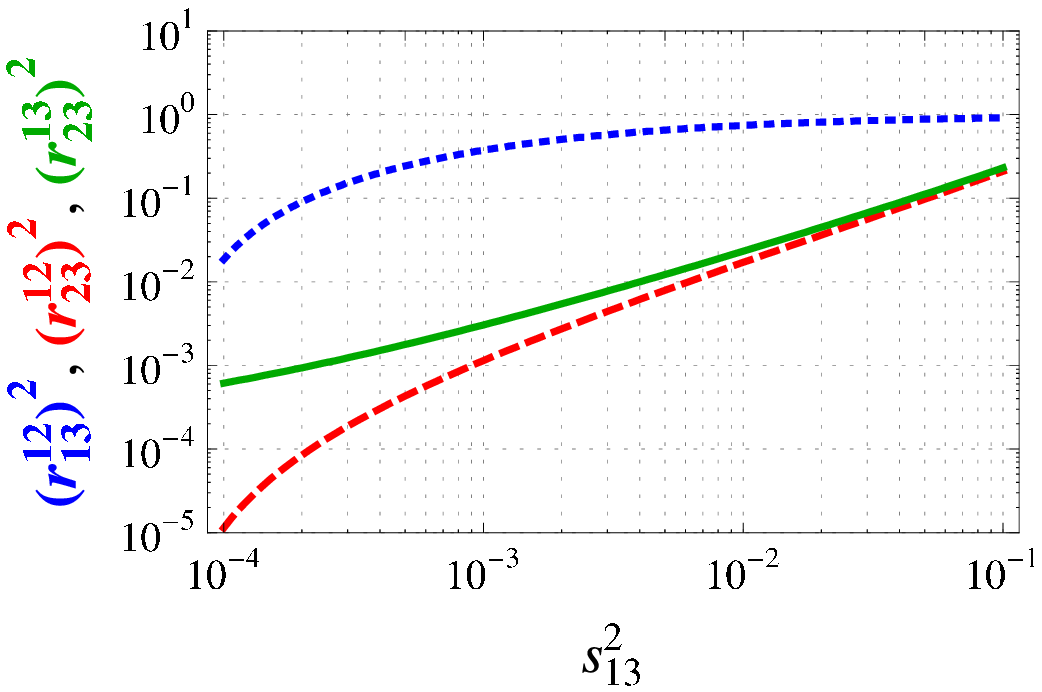} &
\includegraphics[width=0.23\textwidth]{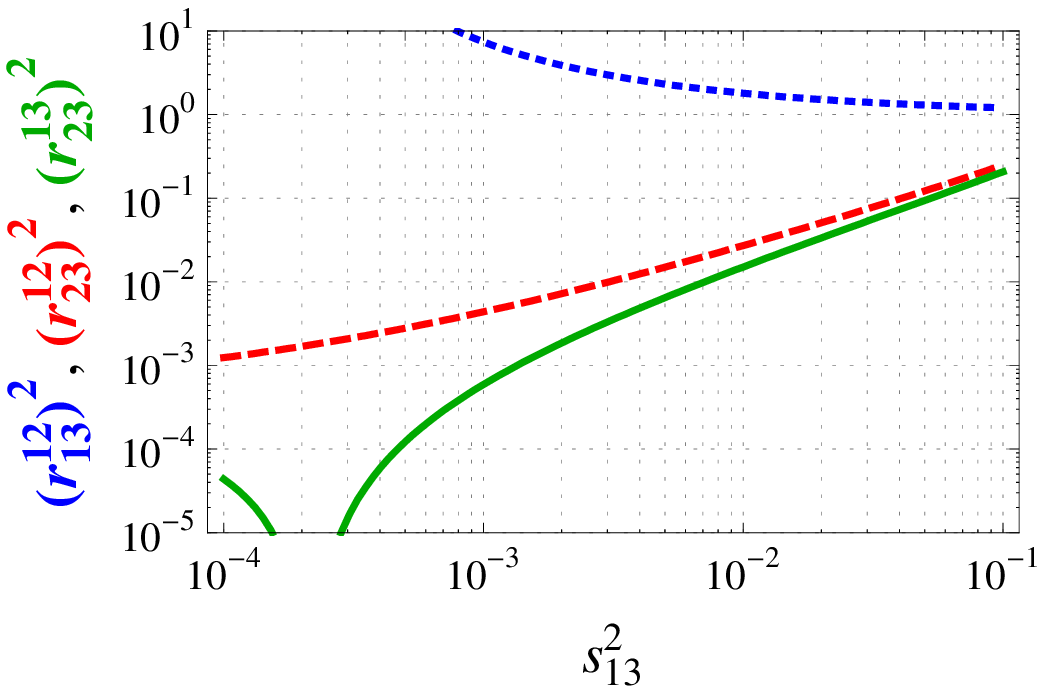} &
\includegraphics[width=0.23\textwidth]{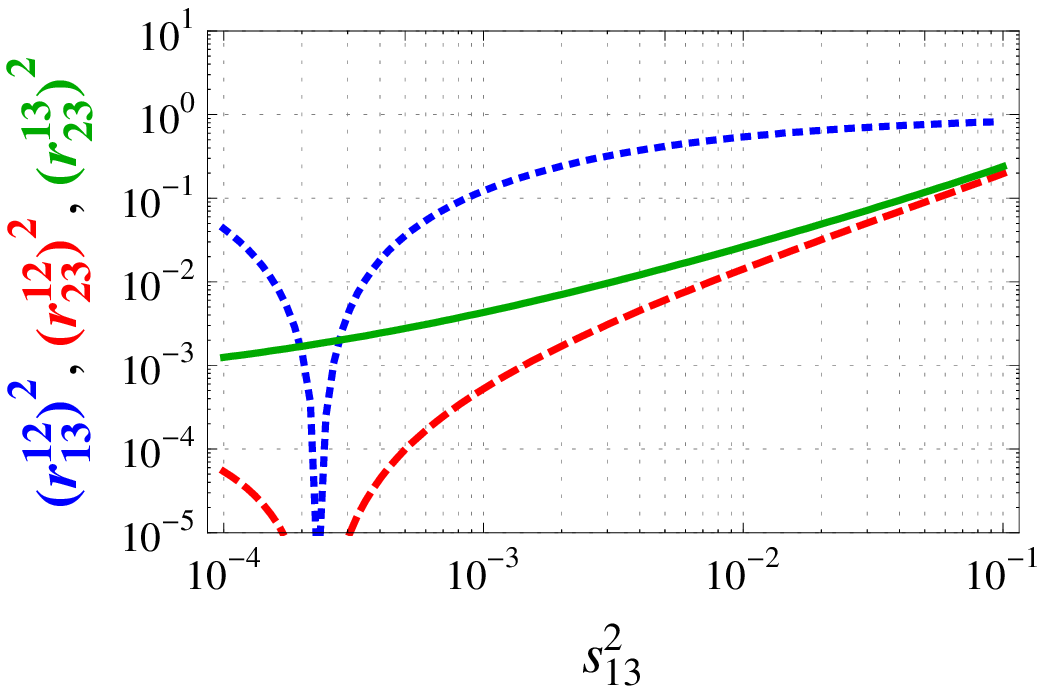} \\\hline
\hspace{0.30cm}%$\vspace{-2cm}
\begin{rotate}{90}{$\qquad\ \ \boldsymbol{\delta=\pi}$}\end{rotate} &
\includegraphics[width=0.23\textwidth]{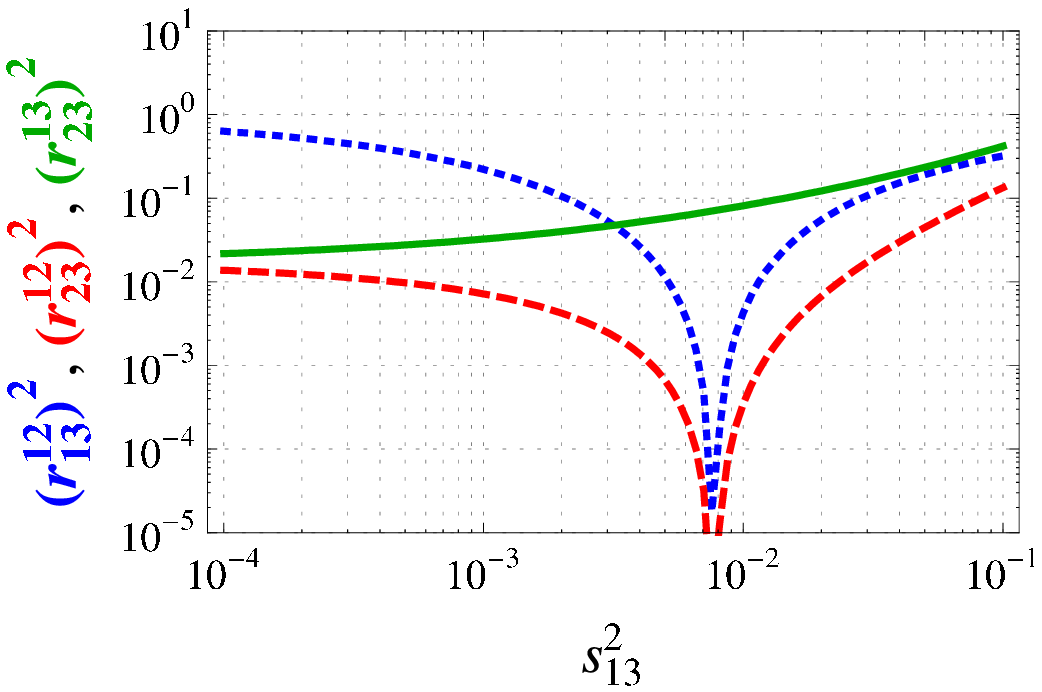}  &
\includegraphics[width=0.23\textwidth]{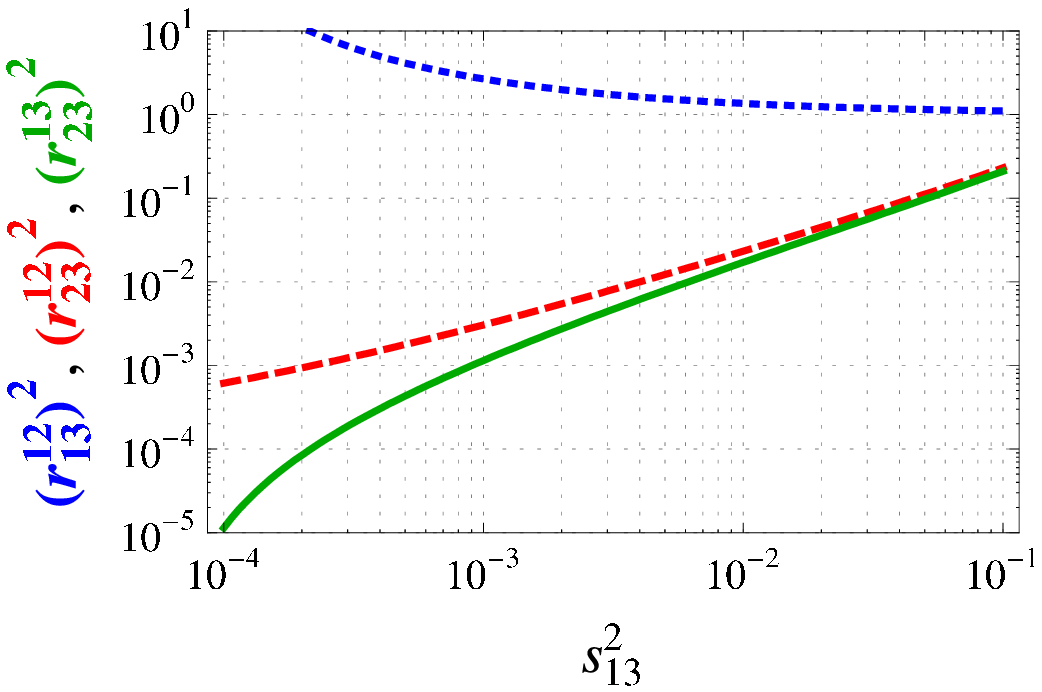} &
\includegraphics[width=0.23\textwidth]{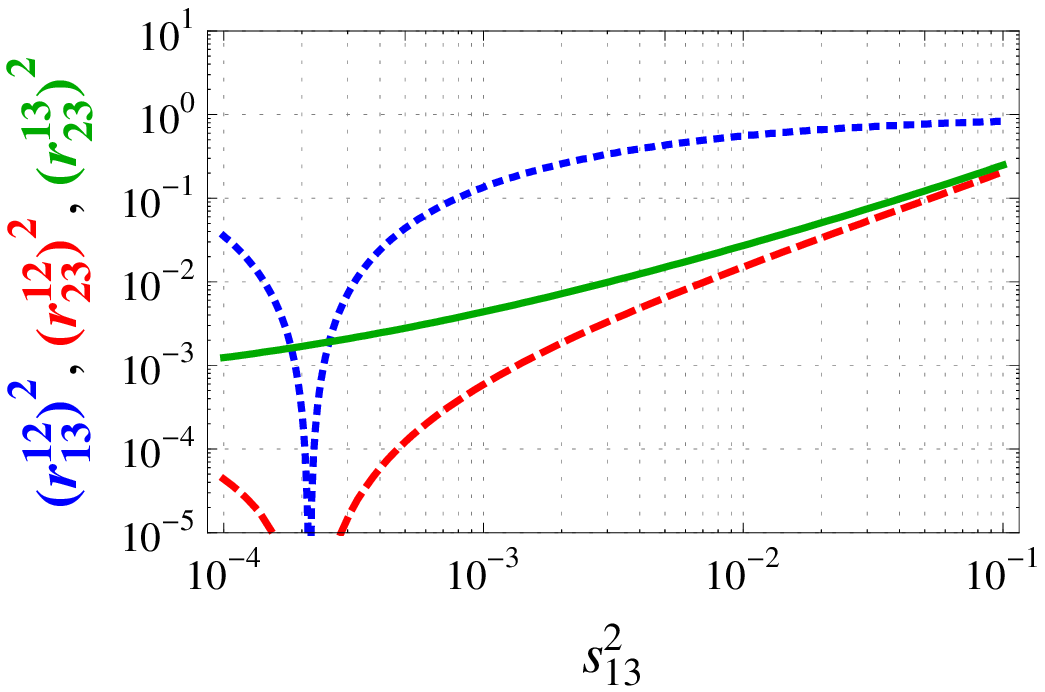} &
\includegraphics[width=0.23\textwidth]{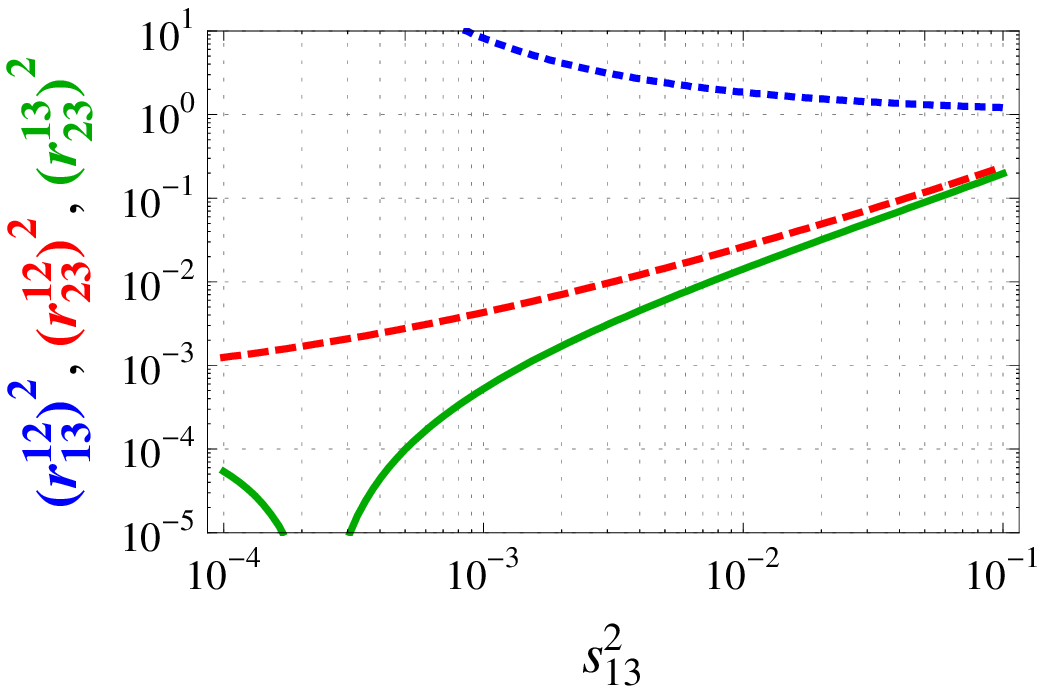} \\\hline
\end{tabular}
\caption{Square ratios $(r^{12}_{13})^2$ (blue line, dotted line), $(r^{12}_{23})^2$ (red line, dashed line) and $(r^{13}_{23})^2$ (green line, full line) versus $s_{13}^2$ for the case of degenerate heavy neutrinos, real $R$ and the rest of the neutrino parameters fixed to their best fit point values. 
Each column correspond to a different neutrino scenario: SNH (first row), SIH (second row), QDNH (third row) and QDIH (forth row). 
Each row correspond to a different value of the Dirac phase: $\delta=0$ (first column) and $\delta=\pi$ (second column).} 
\label{fig:DegNuR-s13}
\end{figure}

\begin{figure}[htb]
\centering
\begin{tabular}{|c|c|c|c|}\hline%\vspace{0.1cm}
%\hspace{0.6cm}\B&
\multicolumn{2}{|c|}{\bf Dominant $\boldsymbol{M_1}$} & \multicolumn{2}{|c|}{\bf Dominant $\boldsymbol{M_2}$} \\\hline
{$\boldsymbol{\delta=0}$} & {$\boldsymbol{\delta=\pi}$} & {$\boldsymbol{\delta=0}$} & {$\boldsymbol{\delta=\pi}$} \\\hline
%
%\vspace{2cm}
%\hspace{0.2cm}
%\begin{rotate}{90}{$\qquad\quad$\bf Dominant $\boldsymbol{M_1}$}\end{rotate} &
\includegraphics[width=0.24\textwidth,clip]{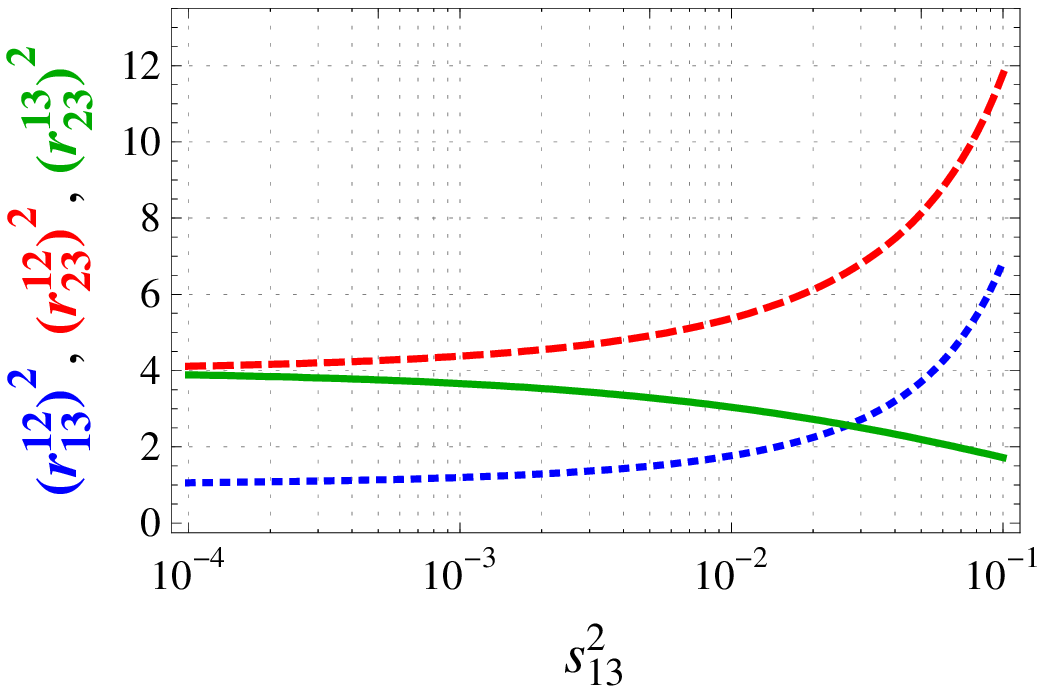}& % 0.3 % 0.38
\includegraphics[width=0.24\textwidth]{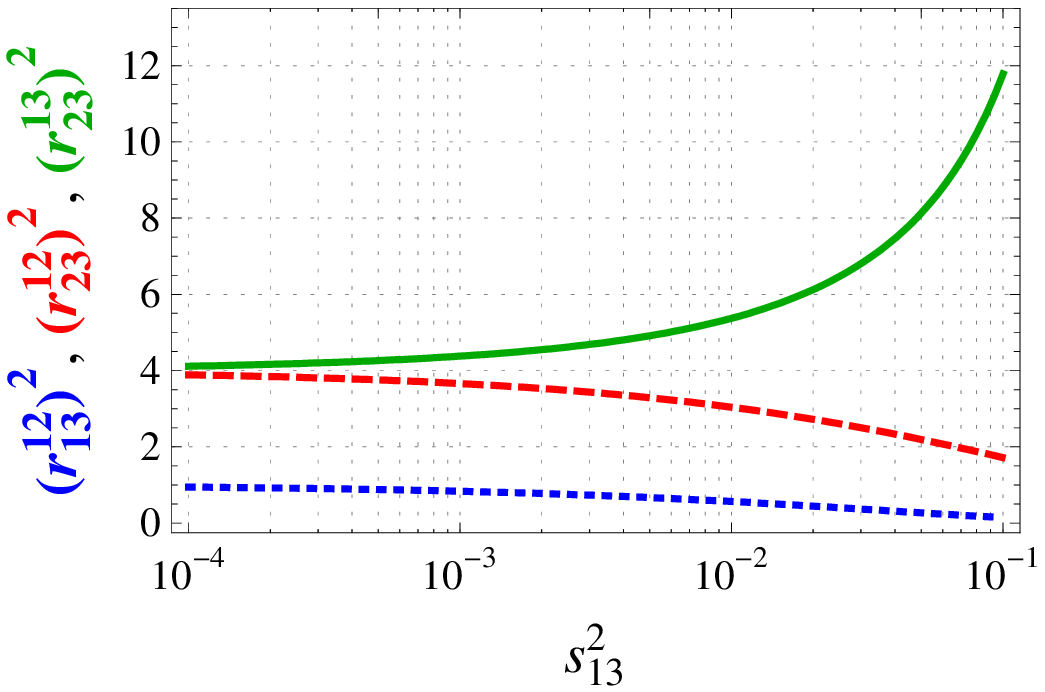} &%\\\hline
%
%\hspace{0.35cm}%$\vspace{-2cm}
%\begin{rotate}{90}{$\qquad\quad$\bf Dominant $\boldsymbol{M_2}$}\end{rotate} 
%&
\includegraphics[width=0.24\textwidth,clip]{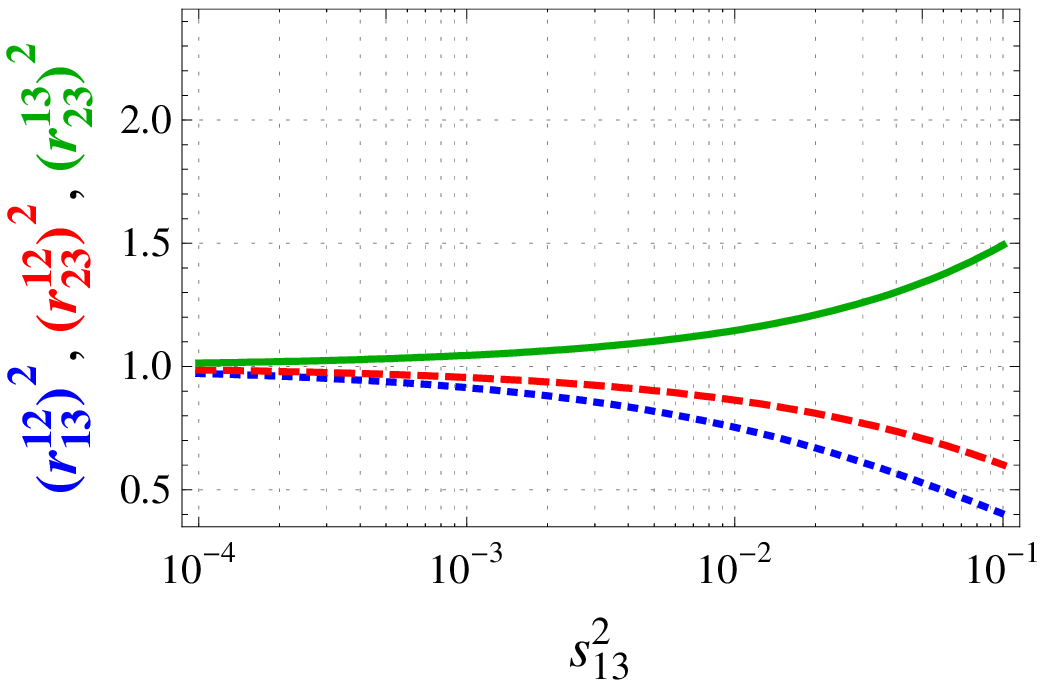}&
\includegraphics[width=0.24\textwidth]{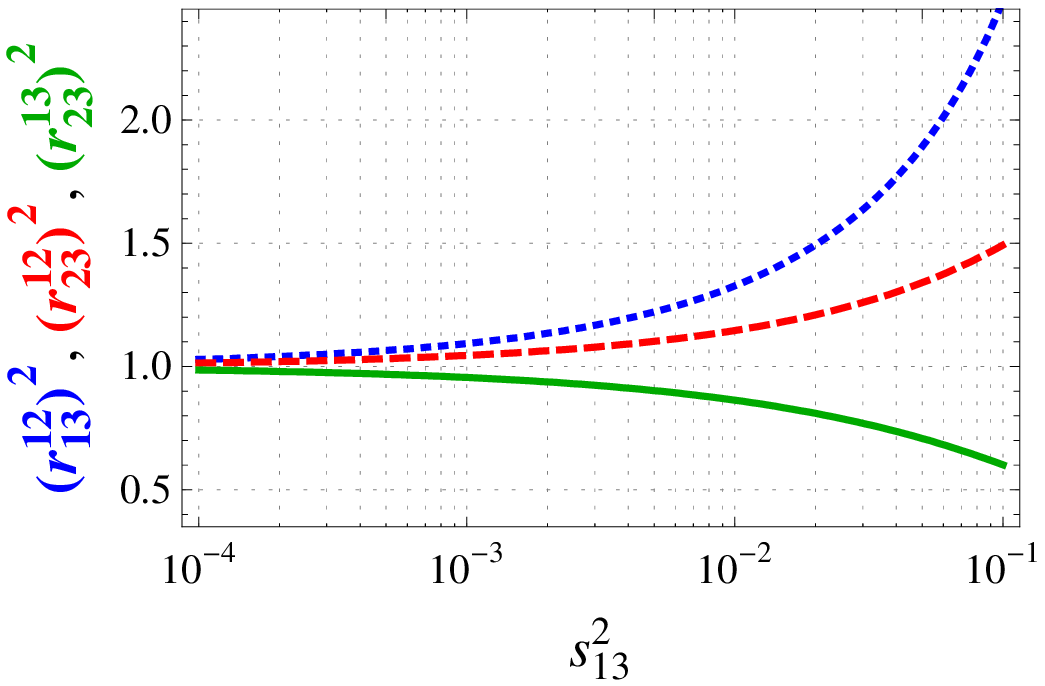} \\\hline
\end{tabular}
\caption{\label{fig:Hier-NuR}Square ratios $(r^{12}_{13})^2$ (blue line, dotted line), $(r^{12}_{23})^2$ (red line, dashed line) and $(r^{13}_{23})^2$ (green line, full line) versus $s_{13}^2$ for the case of strongly hierarchical heavy neutrinos, $R$ being the identity and the rest of the neutrino parameters fixed to their best fit point values. 
First and second panel correspond to $\delta=0$ and $\delta=\pi$, respectively, in the case of dominant $M_1$, while third and fourth panel correspond to $\delta=0$ and $\delta=\pi$, respectively, in the case of dominant $M_2$.} 
\end{figure}

%
%\subsection{Degenerate right-handed neutrinos}
In the case of degenerate right-handed neutrinos and assuming $R$ being real and tribimaximal (TBM) mixing~\cite{Harrison:2002er}, 
\begin{equation}\label{eq:DegNuR-TBM}
r^{13}_{23}  = \frac{2 (m_2-m_1)}{|3 m_3-2 m_2-m_1|}.
\end{equation}
Table~\ref{tab:DegNuR-TBM} shows the form of Eq.~(\ref{eq:DegNuR-TBM}) and its numerical values, for different neutrino scenarios. %: strict normal hierrachy (SNH), strict inverse hierrachy (SIH) and quasidegenerate neutrinos (QD).
\begin{table}[htb]
\begin{center}
\caption{Parameter $r^{13}_{23}$ for the case of degenerate right-handed neutrinos, assuming $R$ being real and TBM mixing. Each column corresponds to a different neutrino scenario: strict normal hierarchical (SNH), strict inverse hierarchical (SIH), quasidegenerete normal hierarchical (QDNH) and quasidegenerete inverse hierarchical (QDIH) neutrinos. First row shows the analytical form of the parameter $r^{13}_{23}$; second row (BFP) shows its value when fixing neutrino mass splittings to their best fit point value and third row shows its value when considering the $3\sigma$ allowed range for the neutrino mass-squared differences. Note that $\alpha \equiv \Ds/|\Da|$ is the ratio of the solar over the atmospheric mass splitting and $\sigma_{\textsc{a}}\equiv\Da/|\Da|$ is the sign of the atmospheric mass splitting.}
\begin{tabular}{|c|c|c|c|c|}\cline{2-5}%\hline
\multicolumn{1}{c|}{} & {\bf SNH} & {\bf SIH} & {\bf QDNH} & {\bf QDIH} \\\cline{2-5}%\hline
\multicolumn{1}{c|}{} & $r^{13}_{23} = \frac{2\sqrt{\alpha}}{3\sqrt{1+\alpha}-2\sqrt{\alpha}}$ & 
$r^{13}_{23} = \frac{2(1-\sqrt{1-\alpha})}{2+\sqrt{1-\alpha}}$ & 
\multicolumn{2}{|c|}{$r^{13}_{23} \simeq \frac{2\alpha}{3\sigma_{\textsc{a}}+\alpha}$} \\\hline
{\bf BFP} & $(r^{13}_{23})^2 = 1.7\times 10^{-2}$ & $(r^{13}_{23})^2 = 1.1\times 10^{-4}$ & $(r^{13}_{23})^2 = 4.4\times 10^{-4}$ & $(r^{13}_{23})^2 = 4.6\times 10^{-4}$ \\\hline
$\boldsymbol{3\sigma}$ & $(r^{13}_{23})^2 \in [0.92,\,3.7]\times 10^{-2}$ & $(r^{13}_{23})^2 \in [0.48,\,3.3]\times 10^{-4}$ & $(r^{13}_{23})^2 \in [1.8,\,12]\times 10^{-4}$ & $(r^{13}_{23})^2 \in [1.9,\,13]\times 10^{-4}$ \\\hline
\end{tabular}
\label{tab:DegNuR-TBM}
\end{center}
\end{table}
%
% Figure removed
%
For the case $s_{13}\ne 0$, Fig.~\ref{fig:DegNuR-s13} shows the squared ratios as a function of $s_{13}^2$, for different neutrino scenarios and for two choices of the Dirac phase $\delta$.
%
%%%%%%%%%%%%%%%%%%%%%%%%%%%%%%%%%%%%%%%%%%%%%%%%%%%%%%%%%%%%%%%%%%%%%%%%%%%%%%%%%%%%%%%%%%%%%%%%%%%%
%\subsection{Right-handed neutrinos strongly hierarchical}
For strongly hierarchical right-handed neutrinos and assuming $R$ being the identity, the parameter $r^{13}_{23}$ does not depend on the light neutrino masses any more, but on their mixing angles and Dirac phase. For the specific case in which $s_{13}=0$, Table~\ref{tab:HierNuR} shows the analytical form of $r^{13}_{23}$ and its numerical values, for different right-handed neutrino dominant mass scenarios.
\begin{table}[htb]
\begin{center}
\caption{Parameter $r^{13}_{23}$ for the case of strongly hierarchical right-handed neutrinos, assuming $R$ being the identity and $s_{13}=0$. Each column corresponds to a different heavy neutrino scenario: if $M_1$ is the heaviest mass eigenvalue (Dominant $M_1$), if $M_2$ is the heaviest mass eigenvalue (Dominant $M_2$) and if $M_3$ is the heaviest mass eigenvalue (Dominant $M_3$). First row shows the analytical form of the parameter $r^{13}_{23}$; second row (BFP) shows its value when fixing neutrino mixing angles to their best fit point value and third row shows its value when considering the $3\sigma$ allowed range for the neutrino mixing angles.}
\begin{tabular}{|c|c|c|c|}\cline{2-4}%\hline
\multicolumn{1}{c|}{} & {\bf Dominant} $\boldsymbol{M_1}$ & {\bf Dominant} $\boldsymbol{M_2}$ & {\bf Dominant} $\boldsymbol{M_3}$ \\\cline{2-4}%\hline
\multicolumn{1}{c|}{} & $r^{13}_{23} = \frac{c_{12}}{s_{12}c_{23}}$ & 
$r^{13}_{23} = \frac{s_{12}}{c_{12}c_{23}}$ & 
$r^{13}_{23} = 0$ \\\hline
{\bf BFP} & $(r^{13}_{23})^2 = 4$ & $(r^{13}_{23})^2 = 1$ & $(r^{13}_{23})^2 = 0$\\\hline
$\boldsymbol{3\sigma}$ & $(r^{13}_{23})^2 \in [2.3,\,8.6]$ & $(r^{13}_{23})^2 \in [0.53,\,2.0]$ & $(r^{13}_{23})^2 = 0$ \\\hline
\end{tabular}
\label{tab:HierNuR}
\end{center}
\end{table}
For the case of nonzero values of $s_{13}$, Fig.~\ref{fig:Hier-NuR} shows the squared ratios as a function of $s_{13}^2$, for different right-handed neutrino dominant mass scenarios. Note that for the scenario of dominant $M_3$, the parameter $r^{13}_{23}$ is proportional to $s_{13}$.

%%%%%%%%%%%%%%%%%%%%%%%%%%%%%%%%%%%%%%%%%%%%%%%%%%%%%%%%%%%%%%%%%%%%%%%%%%%%%%%%%%%%%%%%%%%%%%%%%%%%%%%%%%%%%%%%%
\section{RESULTS}
In order to check the validity of our analytical estimated ratio of stau LFV decays, we have performed a numerical calculation of such LFV decays making use of the program package \textsc{spheno}~\cite{Porod:2003um}. More details %about the numerical procedure 
can be found in Ref.~\cite{Hirsch:2008dy}.

\begin{figure}[htb]
\centering
\begin{tabular}{|c|c|c|c|}\hline%\vspace{0.1cm}
%\hspace{0.6cm}\B 
{\bf Degenerate $\boldsymbol{M_i}$} & {\bf Dominant $\boldsymbol{M_1}$} &  {\bf Dominant $\boldsymbol{M_2}$} & {\bf Dominant $\boldsymbol{M_3}$} \\\hline
\includegraphics[width=0.24\textwidth]{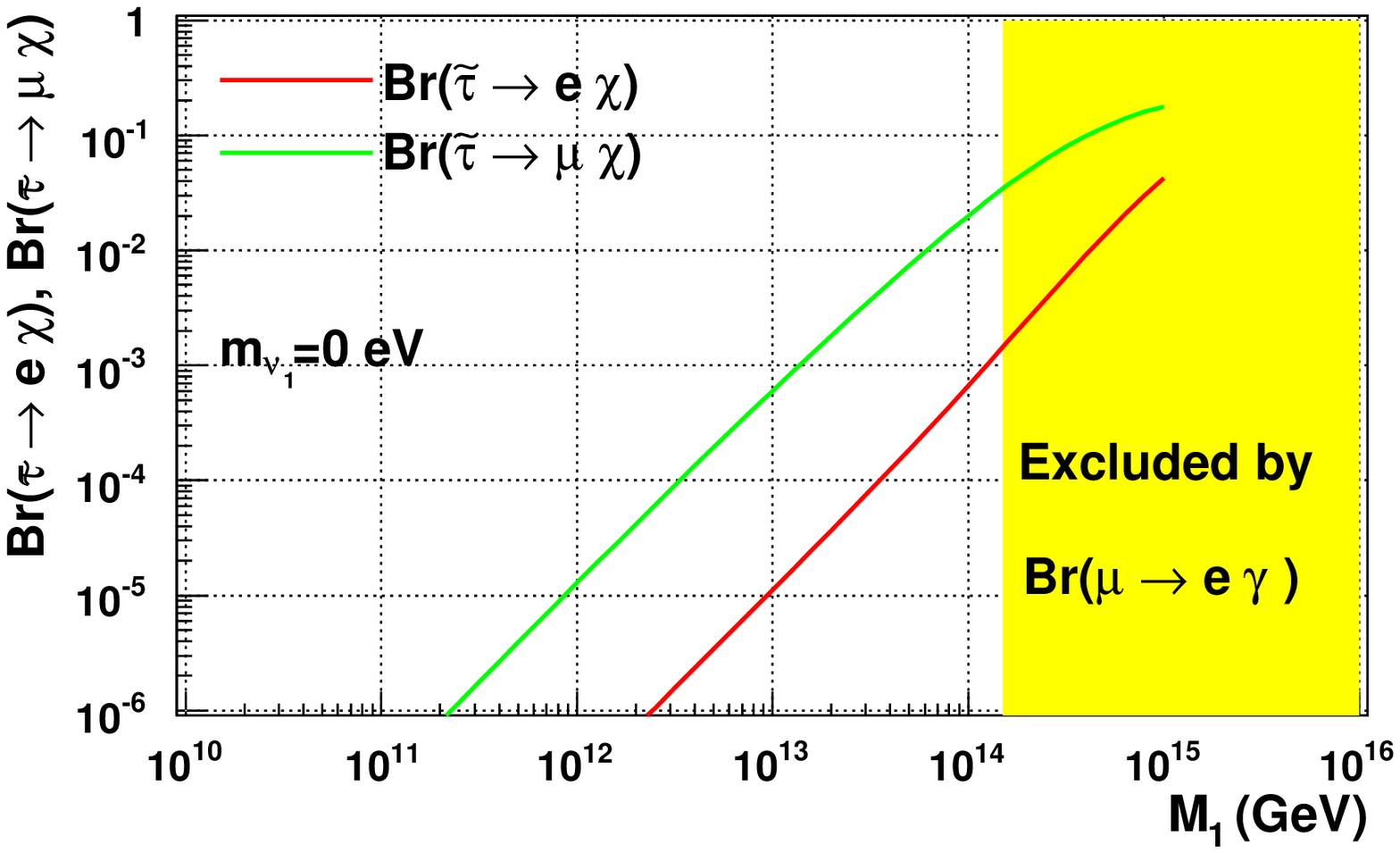} &
\includegraphics[width=0.24\textwidth]{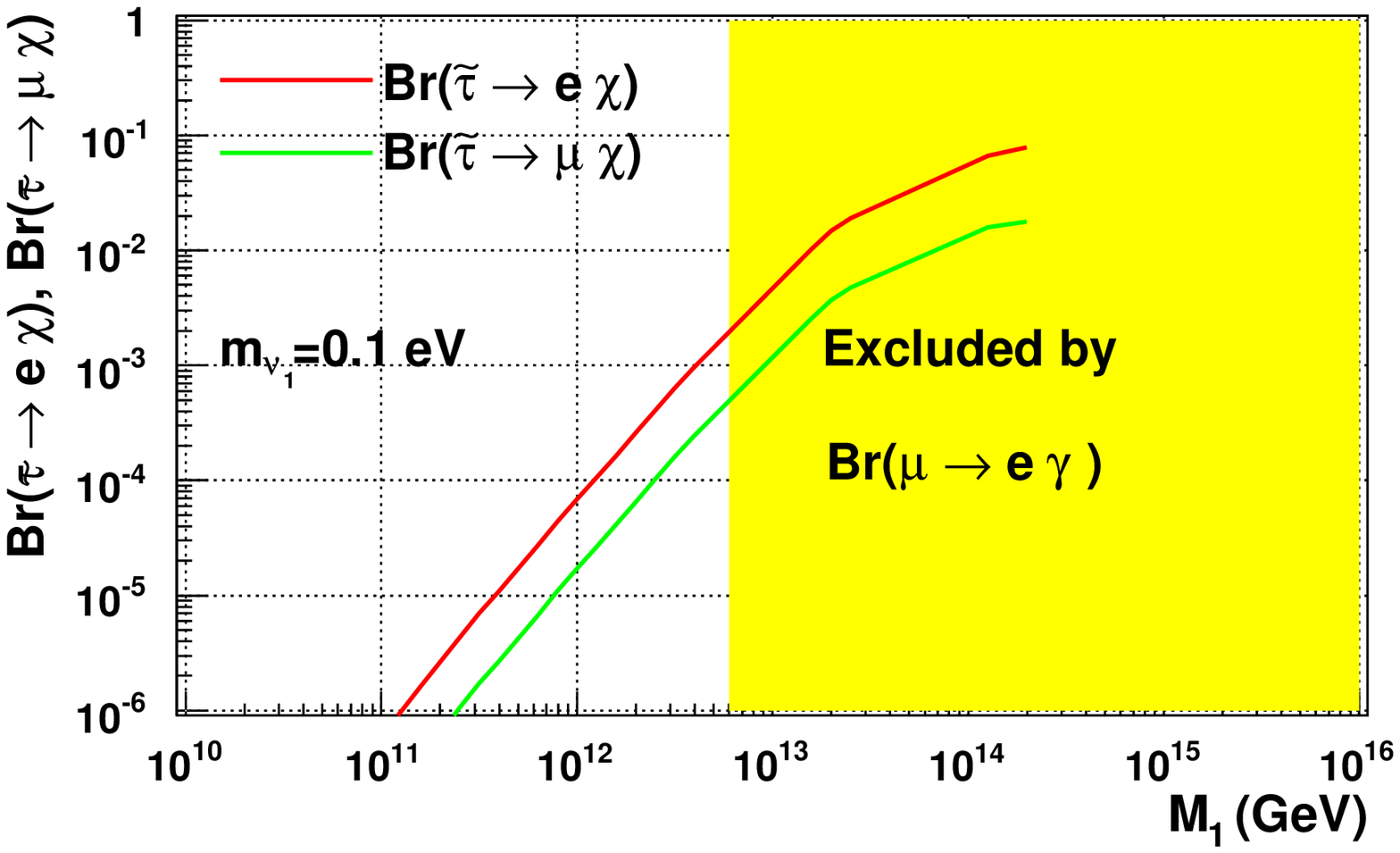} &
\includegraphics[width=0.24\textwidth]{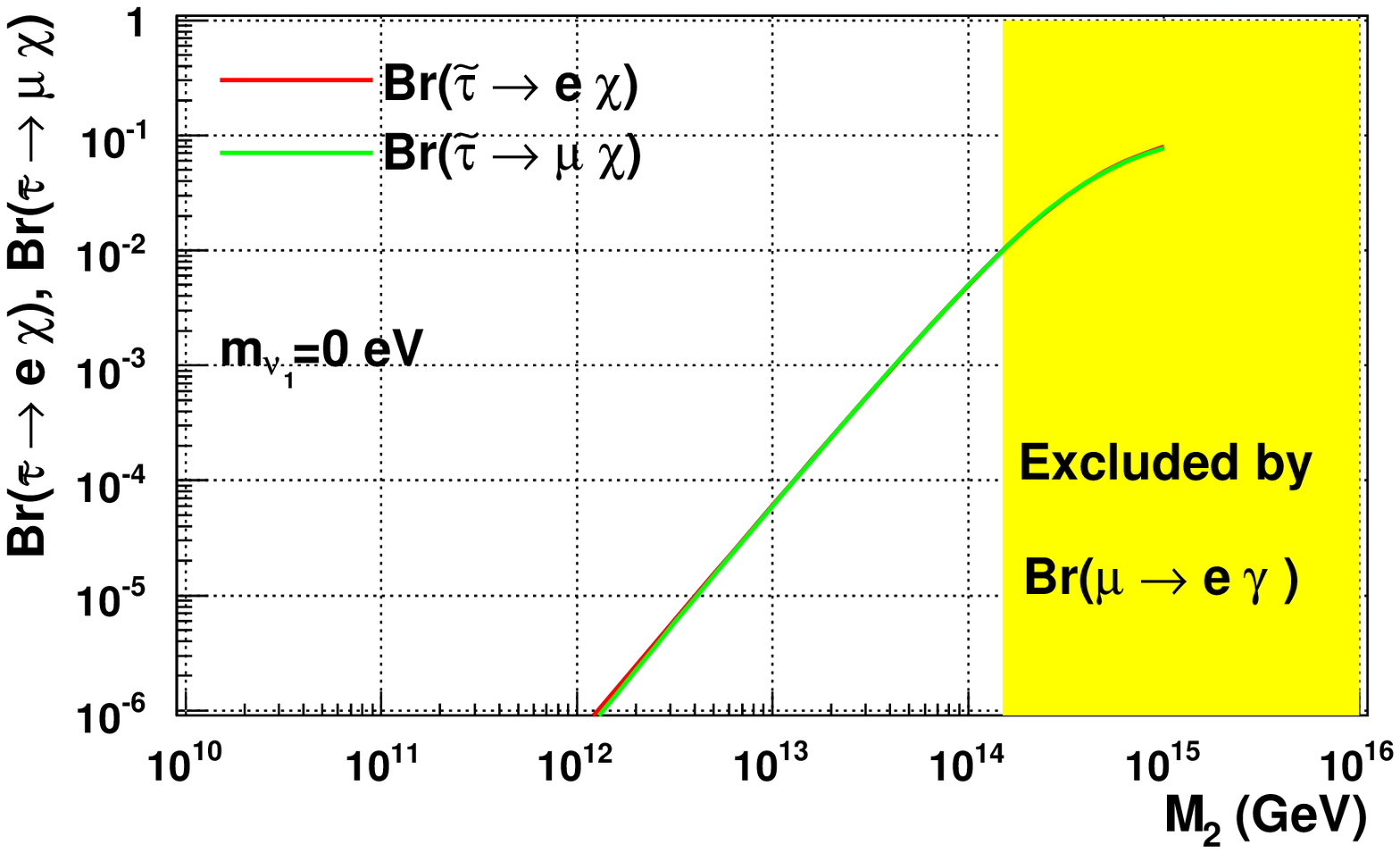} &
\includegraphics[width=0.24\textwidth]{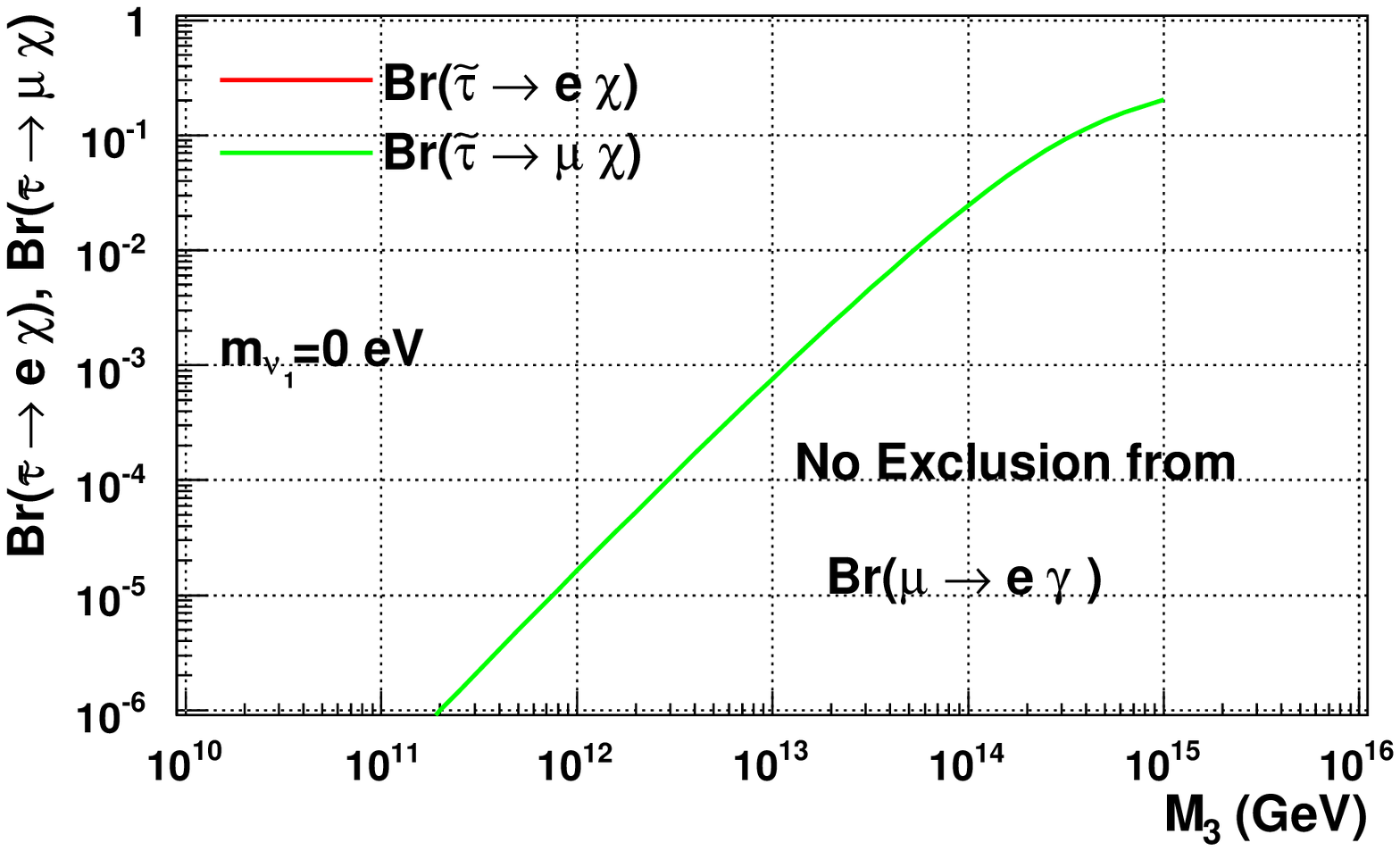} \\\hline
\end{tabular}
\caption{\label{fig:results}Left-stau LFV decays as a function of the right-handed neutrino degenerate common mass $M_R$ (first panel), of the dominant $M_1$ (second panel), of the dominant $M_2$ (third panel) and of the dominant $M_3$ (fourth panel), for the SPS3 standard point, $R$ being the identity, TBM mixing and SNH (except for the second panel, where $m_1=0.1$ eV). In the panels where one $M_i$ is dominant, the other two heavy neutrino masses have been fixed to $10^{10}$ GeV.} 
\end{figure}

%\subsection{Degenerate right-handed neutrinos}
For degenerate right-handed neutrinos, first panel in Fig.~\ref{fig:results} shows the stau LFV decays as a function of the heavy neutrino common mass $M_R$, for mSugra benchmark point SPS3~\cite{Allanach:2002nj}, SNH ($m_1=0$) and TBM mixing.

%\subsection{Right-handed neutrinos strongly hierarchical}
For strongly hierarchical right-handed neutrinos, the other panels in Fig.~\ref{fig:results} show the stau LFV decays as a function of the heavy neutrino dominant mass, which is $M_1$ in the second panel, $M_2$ in the third panel and $M_3$ in the fourth panel. Again, NH and TBM have been assumed.

Note that the ratio of the stau LFV decays follows very accurately the analytical estimate in the region allowed by the upper limit on Br($\mu\to e\gamma$).
%%%%%%%%%%%%%%%%%%%%%%%%%%%%%%%%%%%%%%%%%%%%%%%%%%%%%%%%%%%%%%%%%%%%%%%%%%%%%%%%%%%%%%%%%%%%%%%%%%%%%%%%%%%%%%%%%
\section{CONCLUSION}
Neutrino experimental data show that neutrinos are massive and mix. If its origin is the simplest supersymmetric type-I seesaw mechanism and mSugra boundary conditions hold, then LFV decays are related to neutrino parameters. In particular, we have studied the relation of the ratio of the stau LFV decays with neutrino parameters for different neutrino scenarios.
\begin{acknowledgments}
The author wish to thank his collaborators M.~Hirsch, W.~Porod, J.~C.~Romao and J.~W.~F.~Valle.
The author is supported by {\it Funda\c c\~ao para a Ci\^encia e a
Tecnologia} under the grant SFRH/BPD/30450/2006. 
This work was partially supported by {\it Funda\c c\~ ao para a
Ci\^ encia e a  Tecnologia} through the projects
POCI/81919/2007 and CFTP-FCT UNIT 777,  which are partially funded
through POCTI (FEDER) and by the Marie Curie RTNs MRTN-CT-2006-035505
and MRTN-CT-2004-503369. 
It was also partially supported by Spanish grants FPA2005-01269 and Accion Integrada
HA-2007-0090 (MEC) and by the European Commission network
MRTN-CT-2004-503369 and ILIAS/N6 RII3-CT-2004-506222.
\end{acknowledgments}

%%%%%%%%%%%%%%%%%%%%%%%%%%%%%%%%%%%%%%%%%%%%%%%%%%%%%%%%%%%%%%%%%%%%%%%%%%%%%%%%%%%%%%%%%%%%%%%%%%%%%%%%%%%%%%%%%
%\begin{thebibliography}{9}   % Use for  1-9  references

\end{document}